\documentclass[twocolumn,pra,superscriptaddress,showpacs]{revtex4-1}
\usepackage{amsmath,multirow,graphicx,amssymb,dcolumn,mathrsfs,bm,color,leftidx,array}
\makeatletter

\newcommand{\Rmnum}[1]{\expandafter\@slowromancap\romannumeral #1@}
\makeatother

\begin{document}

\title{coherent population transfer via state-independent quasi-adiabatic dynamics}

\author{Jian Xu}
\email{xujian$_$328@163.com}\affiliation{College of Electronics and Information Engineering, Guangdong Ocean University, Zhanjiang, 524088,
China}

\author{Yan-Xiong Du}
\affiliation{Guangdong Provincial Key Laboratory of Quantum Engineering and Quantum Materials,
SPTE, South China Normal University, Guangzhou 510006, China}

\author{Wei Huang}
\email{huangwei00@tsinghua.org.cn}\affiliation{Guangdong Provincial Key Laboratory of Quantum Engineering and Quantum Materials, 
SPTE, South China Normal University, Guangzhou 510006, China}

\date{\today}

\begin{abstract}

High-fidelity and robust coherent population transfer is a major challenge in coherent quantum control. Different from the well known adiabatic condition, we present a rigorous adiabatic condition that is inspired by the idea of the Landau-Zener tunneling. Based on this, we propose a coherent population transfer approach, which just needs only one control parameter and depends on the eigenvalues of the systems. Compared to other approaches, such as fast quasiadiabatic dynamics, shortcut to adiabatic passage, we numerically demonstrate that our approach can provide a more high-fidelity and more robustness coherent population transfer without affecting the speed. In short, our approach opens a new way to further increase the fidelity and the robustness of coherent population transfer. Moreover, it may be generalized to complex quantum systems where the exact expressions of eigenstates are difficult to obtain or the paremeters of systems are difficult to simultaneously drive.

\end{abstract}


\maketitle

\section{Introduction}

Coherent population transfer of the quantum state is a major scientific and technological challenge in various areas of physics\cite{Walmsley}. In most applications, the basic requirement of the coherent quantum control is to reach a given target state with the high-fidelity as fast as possible, with sufficiently high fidelity allowed by available resources and experimental constraints. Rabi oscillation can reach very high fidelity but it is so sensitive to the parameter  error or the environmental noise. To overcome this, adiabatic approaches such as rapid adiabatic passage (RAP) and stimulated Raman adiabatic passage (STIRAP)\cite{Kral,Saffman,Vitanov} are proposed to robustly realize high-fidelity state transfer but so time-consuming. Lately the quantum operations whose the fidelity can approach the threshold of quantum computing\cite{Knill,Ladd} have realized in trapped ion\cite{Ballance,Gaebler}. Then, how to speed-up the coherent population transfer while keeping the fidelity and robustness of the operation becomes a challenging question.

To reach this goal, several protocols called "shortcut to adiabaticity" (STA) have been recently proposed theoretically\cite{Demirpla,Berry,Chen,Torosov,Steane,Baksic,Li,H}. And the relevant experiments in different areas have great developments, for instance optical lattice\cite{Bason}, cold atom\cite{Du}, trapped-ion\cite{An,Schafer}, NV-center\cite{Zhou} and Fermi gas\cite{Deng1,Deng2}. But they are not always easy to implement in multilevel  systems, because of various parameters needed to control. So another set of approaches called quasiadiabatic dynamics (QUAD) that only one single control parameter needs to engineer have proposed\cite{Roland,Quan,Martinez-Garaot1}. Recently, the corresponding theories have demonstrated in different experiments, such as trapped-ion\cite{Bowler,Richerme}, ultracold atom\cite{Martinez-Garaot2} and optical waveguides\cite{Martinez-Garaot3,Chung1,Chung2}. But above STA and QUAD theories must depend on the exact expressions of the eigenstates, so these approaches are hard to generalize to complex systems. Recently, a STA approach without dependent of the expressions of adiabatic eigenstates has been proposed\cite{Ran}. So is it possible to build a QUAD approach that just needs one control parameter and be independent of the expressions of eigenstates? Further, due to not having the problems of the match of control parameters and facing to the sensitivity of eigenstates to control parameters, is it more robust against the control parameter variations?

In this paper, we propose the state-independent quasiadiabatic dynamics (SIQUAD) that only depends on one single control parameter and just relates to the eigenvalues of the system. We first propose a rigorous adiabatic condition in the Landau-Zener Hamiltonian and then deduce the relevant population transfer approach. Based on this, we apply our approach to the two- and the three-level systems and find that our scheme has higher fidelity and more robustness than other approaches without affecting the speed. Thus, our work opens up a new possibility for realizing a fast, high-fidelity and robust quantum state transfer in a wider class of systems.

The paper is organized as follows. In Sec. II, we propose the rigorous adiabatic condition in the Landau-Zener model to present SIQUAD. Sec. III we contrast SIQUAD with the other approaches in the two- and the three-level systems. Finally, a brief conclusion are given in Sec. IV.

\section{state-independent quasiadiabatic approach}

In this section, we first rewrite the adiabatic condition in the Landau-Zener tunneling Hamiltonian to deduce SIQUAD. In the simplest model we assume that the adiabatic process involves a passage through at least one avoided crossing by a monotonous change of one parameter. Although the system is multilevel in general, only the two quasi-crossing levels($E_{\pm}(t)$) in the instantaneous basis($|\phi_{\pm}(t)\rangle$) are considered under the adiabatic condition\cite{Schiff,Albash}. Considering the two-level Landau-Zener model\cite{Landau,Zener,Shevchenko}($\hbar=1$):
\begin{equation}
\label{H_lz}
\hat{H}_{LZ}=\frac{1}{2}\left(\begin{matrix}\delta(t)&\Omega^*\\\Omega&-\delta(t)\end{matrix}\right),
\end{equation}
with $\delta(t)$ being a monotonous change parameter and $\Omega$ being a constant. The corresponding eigenvalues and the eigenstates are $E_{\pm}(t)=\pm \frac{1}{2} \sqrt{\delta(t)^2+\Omega^2}$ and 
\begin{equation}
\label{es}
\begin{split}
|\phi_+(t)\rangle&=\left(\begin{matrix}\sin(\theta/2)\\ \cos(\theta/2)\end{matrix}\right) \\
|\phi_-(t)\rangle&=\left(\begin{matrix}\cos(\theta/2)\\ -\sin(\theta/2)\end{matrix}\right),
\end{split}
\end{equation}
with $\theta=\arccos(-\delta(t)/\sqrt{\delta(t)^2+\Omega^2})$. 

Then we define $V=(|\phi_+(t)\rangle,|\phi_-(t)\rangle)$ and rewrite the Hamiltonian in the rotating frame:
\begin{equation}
\begin{split}
\label{H2}
&\tilde{H}=
i\frac{dV^{\dag}}{dt}V+V^{\dag}HV
\\&=\frac{1}{2} \begin{pmatrix} E_+(t) & i \partial_t\theta  \\ -i \partial_t\theta & E_-(t)  \end{pmatrix}.
\end{split}
\end{equation}
Mathematically, the adiabatic evolution requires the off-diagonal elements of the Hamiltonian in Eq.(\ref{H2}) to be negligible relative to the diagonal ones, so we have the standard condition for adiabatic evolution\cite{Vitanov}:
\begin{equation}
\label{ac}
s=\frac{|\partial_t\theta|}{\sqrt{\delta(t)^2+\Omega^2}}=\frac{1}{2} \frac{\partial_t \delta(t)\sqrt{\frac{\Omega^2}{\delta(t)^2+\Omega^2}} }{\delta(t)^2+\Omega^2}\ll1,
\end{equation}
which is used as the foundation of other approaches. Comparing to the Landau-Zener tunneling formulation, $s$ is the exponent part of the tunneling formulation except for $\sqrt{\Omega^2/[\delta(t)^2+\Omega^2]}$. Here, due to this term is less than or equal to 1, we omit it to get a more rigorous adiabatic condition:
\begin{equation}
\label{ac2}
s'=\frac{1}{2} \frac{\partial_t \delta(t)}{\delta(t)^2+\Omega^2}\ll 1.
\end{equation}
The above equation can be solved with $\delta(T/2)=0$:
\begin{equation}
\label{delta1}
\delta(t)=\Omega \tan[s' (2t-T)\Omega].
\end{equation}
Considering $\delta(0)=-\delta(T)=-\delta_m (\delta_m\gg\Omega)$, Eq.(\ref{delta1}) can be rewritten as $s'=\arctan(\delta_m/\Omega)]/(T \Omega)$. Based on this condition, we can further rewrite Eq.(\ref{delta1}) as:
\begin{equation}
\label{delta2}
\delta(t)=\Omega \tan[(\frac{2t}{T}-1)\arctan(\delta_m/\Omega)].
\end{equation}

In a specific system, $\delta(t)$ is generally depend on an external parameter, e.g. the detuning of the laser in cold atoms or the energy of the superconducting qubit, and $\Omega$ is equal to the energy gap of the system when $\delta(t)=0$. So this Hamiltonian is easy to be realized in many different systems. On the other hand, we expect our approach has stronger robustness in coherent population transfer because of our more rigorous adiabatic condition in Eq.(\ref{ac2}).

\section{applications}

\begin{figure}[tbp] 
\includegraphics[width=7.5cm]{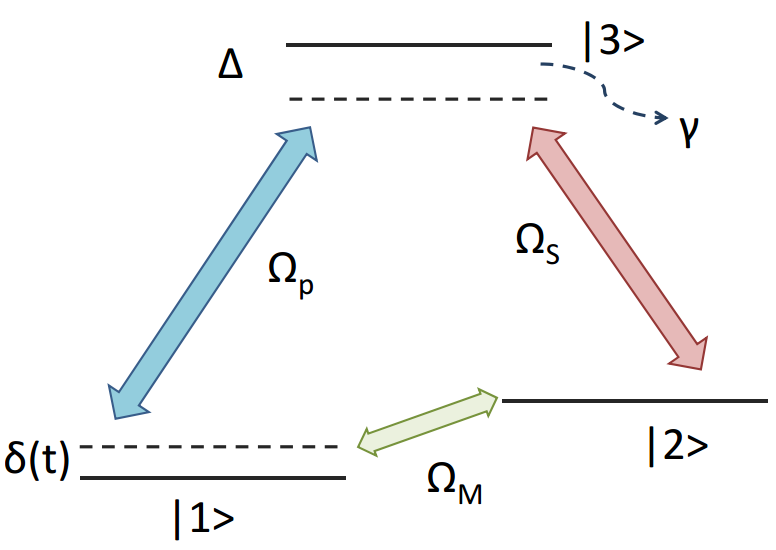}
 \caption{(Color online) Enengy level secheme with the Raman laser field $\Omega_{p,S}=\Omega_0=2\pi\times5MHz$, the microwave field $\Omega_M=2\pi\times150kHz$, the one-photon detuning of the Raman laser field $\Delta=2\pi\times10GHz$, and the value ranges of the effective detuing of the microwave field or the two-photon detuning of the Raman laser field $\delta$ being [$-\delta_m, \delta_m$], with $\delta_m=2\pi\times10MHz$. 
}\label{fig1}
\end{figure}

In this section, we show the experimentally transfer scheme in the two- and the three-level systems. Consider a three-level system with states $\{|1\rangle,|2\rangle,|3\rangle\}$, the interaction Hamiltonian within the rotating wave approximation can be written as($\hbar=1$)
 \begin{equation}
\label{H_Lambda}
\hat{H}_{\Lambda}=\frac{1}{2}\left(\begin{matrix}2\delta(t) & \Omega_M^* & \Omega_p^*\\ \Omega_M & 0& \Omega_S^*\\ \Omega_p & \Omega_S & 2\Delta-2i\gamma\end{matrix}\right), 
\end{equation}
where $\Omega_p$ and $\Omega_S$ are the Rabi frequencies of pump and Stokes fields, $\Omega_M$ is the Rabi frequencies of microwave field and $\gamma$ is the spontaneous emission of the excited state $|3\rangle$. Here $\Delta$ is the single-photon detuning, $\delta(t)$ is the two-photon detuning in the three-level system or the effective detuning between $\Omega_M$ and the two-level system, as shown in Fig.(\ref{fig1}). If we consider the spontaneous emission $\gamma$, our scheme can be demonstrated by the recent experiment of cold atoms\cite{Du}, where the laser-atom three-level coupling scheme are presented, and two ground states are $|1\rangle=|F=1,m_F=0\rangle$, $|2\rangle=|F=2, m_F=0\rangle$; the excited state is $|3\rangle=5^2P_{3/2}$. If $\gamma$ is neglected, one can select the experiment of superconducting transmon qubit\cite{Xu}, where $\{|1\rangle,|2\rangle,|3\rangle\}$ denote the three lowest energy levels of the transmon qubit and $|3\rangle$ is an auxiliary state and remains unoccupied before and after the operation.

\subsection{two-level systems}

When $\Omega_{p,S}=0$, the Hamiltonian in Eq.(\ref{H_Lambda}) is simplified to :
\begin{equation}
\label{H_two}
\hat{H}_{two}=\frac{1}{2}\left(\begin{matrix}2\delta(t)&\Omega_M^*\\\Omega_M&0\end{matrix}\right),
\end{equation}
with the energy gap without $\delta(t)$ being $\Omega_M$ and the detuning of the Rabi frequency being $\delta(t)$. Considering a population inversion with Eq.(\ref{H_two}), suppose the bare states are $|1\rangle=\begin{pmatrix}1,0\end{pmatrix}^{T}$, $|2\rangle=\begin{pmatrix}0,1\end{pmatrix}^{T}$. For $\delta(t)=0$, the $\pi$ pulse Rabi oscillation occurs and the $\pi$ pulse operation time $\tau_\pi\equiv\pi/\Omega_M\approx 3.33\mu s$. When one drives $\delta(t)$ from $\delta(0)=-\delta_m$ to $\delta(T)=\delta_m$, for example our approach (SIQUAD) and fast quasiadiabatic approach (FAQUAD), the eigenstate is drived from $|\phi_-(0)\rangle=\begin{pmatrix}1,0\end{pmatrix}^{T}$ to $|\phi_-(T)\rangle=\begin{pmatrix}0,1\end{pmatrix}^{T}$. 

\begin{figure}[tbp] 
\includegraphics[width=8.7cm]{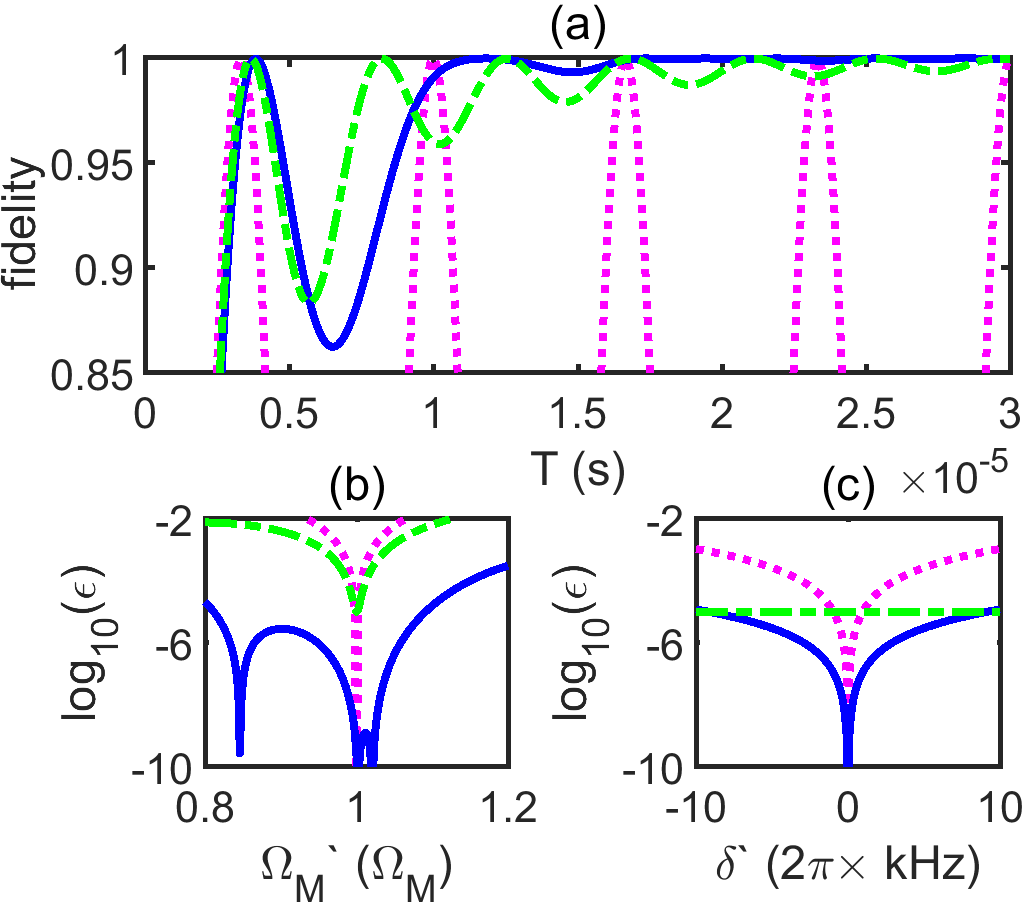}
\caption{(Color online) (a)The transfer fidelity as the function of the operation time in $\pi$ pulse (magenta dotted line), FAQUAD (green dash-dotted line) and SIQUAD (blue solids line). The transfer error $\epsilon$ as the function of (b) the Rabi frequency variations $\Omega_M'$ and (c) the detuning errors of Rabi oscillation (magenta dotted line), FAQUAD (green dashed line) and SIQUAD (blue solids line).}\label{fig2}
\end{figure}

Here, we compare SIQUAD to $\pi$ pulse and FAQUAD in Fig.\ref{fig2}. It is easy to find that there are periodicity for all three approaches. The numerical result shows that SIQUAD and FAQUAD can emerge their advantages when the operation time $T$ is larger than 3 times $\tau_\pi$. Unlike $\pi$ pulse, which is a flat-pulse, SIQUAD and FAQUAD maxima are more stable as operation time $t$ increases. Moreover, fig.\ref{fig2}(a) shows that SIQUAD has a more stability than that of FAQUAD. So further we discuss the robustness against the control parameter variations. Fig.\ref{fig2}(b) shows that without affecting the fidelity SIQUAD ($T\approx5.83\tau_\pi$) is more robust than $\pi$ pulse ($t=\tau_\pi$) and FAQUAD ($t\approx6.33\tau_\pi$), when the variations of laser intensity is induced, described by $\Omega_M'$. Further, Fig.\ref{fig2}(c) demonstrates SIQUAD has more fidelity and robustness against the detuning error, which is denoted by $\delta'$. In short, SIQUAD has a more stability and robustness than the other approaches in two-level systems, while maintaining the fidelity and the speed. Note that all the advantage of SIQUAD over FAQUAD is due to the rigorous adiabatic condition in Eq.(\ref{ac2}). The standard adiabatic condition of FAQUAD is also widely applied in a great variety of techniques for transferring population between two discrete quantum states by coupling them with two radiation fields via an intermediate state, e.g. SITRAP. So now we turn to compare the performance of SIQUAD to that of other approaches in three-level systems.

\begin{figure}[tbp] 
\includegraphics[width=8.7cm]{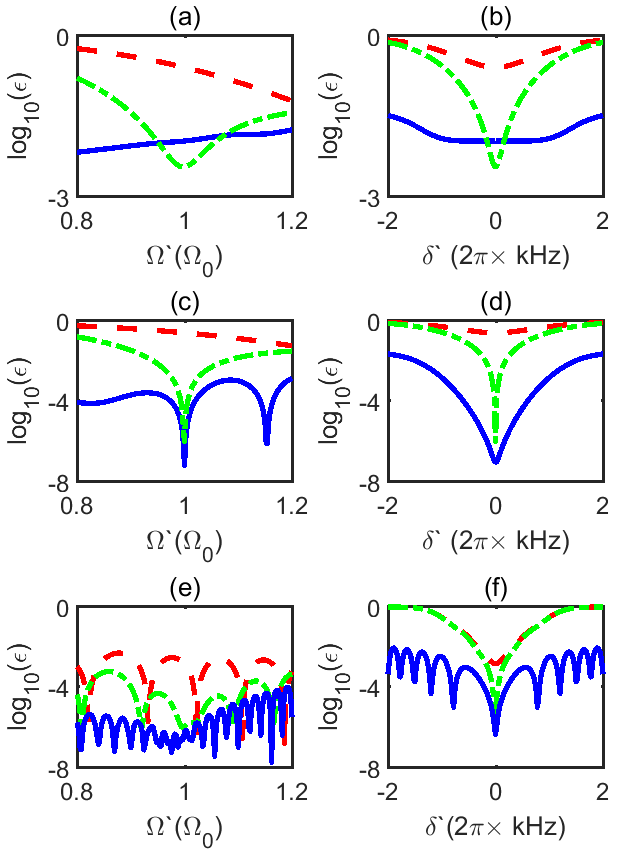}
 \caption{(Color online) 
The transfer error $\epsilon$ as the functions of the Rabi frequency variation $\Omega'$ and the effective detuning eroors $\delta'$ for (a), (b) $\gamma=2\pi\times5.6MHz, T=2.85ms$; (c), (d) $\gamma=0, T=2.85ms$; (e), (f) $\gamma=0, T=21.12ms$ of STIRAP (red dashed line), STIRSAP (green dash-dotted line) and SIQUAD (blue solids line), respectively.}\label{fig3}
\end{figure}

\subsection{three-level systems}

In contrast to resonant $\pi$ pulse, STIRAP is notable because it can immune against loss through spontaneous emission from the intermediate state and is robust against small variations of experimental conditions\cite{Vitanov}. As compared to the case of one-photon resonance, STIRSAP at large one-photon detuning shows the advantages include the avoiding the spontaneous emission from the intermediate state and the more robustness against systematic errors\cite{Li}. Therefore we discuss here the robustness against systematic errors about SIQUAD, STIRAP and STISRAP at large one-photon detuning in three-level systems. When $\Omega_M=0$, the Hamiltonian of Eq.(\ref{H_Lambda}) becomes a common three-level $\Lambda$ Hamiltonian:
 \begin{equation}
\label{H_Lambda2}
\hat{H}_{\Lambda}=\frac{1}{2}\left(\begin{matrix}2\delta(t) & 0 & \Omega_p^*(t)\\ 0 & 0& \Omega_S^*(t)\\ \Omega_p(t) & \Omega_S(t) & 2\Delta-2i\gamma\end{matrix}\right),
\end{equation}
with the peak value of Raman laser field $\Omega_p(t),\Omega_S(t)$ for STIRAP and STIRSAP being $\Omega_0$, the constant value of Raman laser field for SIQUAD also being $\Omega_0$, the absolute value of the lowest two level energy gap without $\delta$ being $\Omega=\sqrt{\Delta^2+\Omega_0^2}-\Delta$ and the corresponding $\pi$ pulse time being $T_0\equiv2\pi\Delta/\Omega^2\approx0.62ms$. As described above, the bare states are $|1\rangle=\begin{pmatrix}1,0,0\end{pmatrix}^{T}$, $|2\rangle=\begin{pmatrix}0,1,0\end{pmatrix}^{T}$ and $|3\rangle=\begin{pmatrix}0,0,1\end{pmatrix}^{T}$ and the population transfer $|1\rangle\leftrightarrow|2\rangle$ can be realized through implementing the two photon detuning $\delta(t)$ in the condition of $\Delta\gg\delta_m\gg\Omega$.

Here, we discuss the robustness of SIQUAD, STIRAP and STISRAP with respect to the Rabi frequency variation $\Omega'$ and the  two photon detuning error $\delta'$, as shown in fig.\ref{fig3}. Fig.\ref{fig3}(a) and \ref{fig3}(b) show the results of SIQUAD, STIRAP and STISRAP at a short operation time ($T=2.85ms\approx4.6T_0$) in the presence of the spontaneous emission $\gamma$ from $|3\rangle$. One can find that SIQUAD has a more robustness than those of STIRAP and STISRAP although all approaches have poor fidelity ($<0.999$). In fig.\ref{fig3}(c) and \ref{fig3}(d), the results in absence of $\gamma$ demonstrate that the performances of SIQUAD and STISRAP have been improved dramatically and SIQUAD precedes STISRAP in fidelity and robustness.  Considering our large one-photon detuning condition $\Delta=2000\Omega_0$, which is much larger than the condition of the experiment\cite{Du}, it means that the spontaneous emission from the intermediate state is hard to suppress perfectly. Further, fig.\ref{fig3}(e) and \ref{fig3}(f) discuss the results at a long operation time ($T=21.12ms\approx34T_0$) because the longer operation time traditionally can improve the performance of STIRAP and STISRAP. Note that the fidelity and the robustness of STIRAP and STISRAP are obviously improved and still poorer than that of SIQUAD, while its the fidelity and the robustness are slightly improved. In short, SIQUAD has a more robustness than those of STIRAP and STISRAP at the same speed and fidelity and can achieve the best performance within a shorter time.

\section{discussion and conclusion}

Here, we summarize the advantages of SIQUAD: (1) it is more robust against the systematic errors than the other approaches without affecting the speed and the fidelity due to a more rigorous adiabatic condition; (2) it can work well in two- and three-level systems. So if the spontaneous emission from the intermediate state in three-level system cannot be perfectly immune against, SIQUAD in two-level case provide a candidate approach to realize a high fidelity and strong robustness coherence population transfer. (3) it just need to control one control parameter and only depend on the information about the eigenvalues of the system, so it has less systematic error sources, e.g. separation time between $\Omega_p$ and $\Omega_S$ in STIRAP and it is easy to generalized to a complex system where there are many systematic parameters.

In summary, we first have theoretically proposed SIQUAD through a rigorous adiabatic condition. Then we apply it to two- and three-level systems to demonstrate that without affecting the speed and the fidelity our approach has more robustness against various systematic errors than the other approaches, which are based on the standard adiabatic condition. Further, note that SIQUAD only depend on a single control parameter and the information about the energy gap, so it is easy to generalize to complex systems, e.g the system where the exact expressions of adiabatic eigenstates are difficult to obtain or many parameters are difficult to simultaneously drive. So our work is both physically transparent and experimental flexible and can to a wide variety of realistic situations.

\acknowledgements

We thank zhi-ming zhang, Feng Mei and Jian-Qi Zhang for helpful discussions. This work was supported by the National Natural Science Foundation of China (11704080, 11704131), the Natural Science Foundation of Guangdong province (2017A030307023, 2016A030310462) and the Project of Enhancing School With Innovation of Guangdong Ocean University (GDOU2017052602).

\end{document}